\documentclass[aps,
		prd,
		reprint,
        superscriptaddress,
        shortbibliography,
		  nofootinbib,
		floatfix,
		notitlepage,
        onecolumn
		]{revtex4-1}
\pdfoutput=1

\usepackage{upgreek}
\usepackage{tikz-feynman}

\usepackage{amsmath}
\usepackage{amssymb}
\usepackage{slashed}
\usepackage{color}
\usepackage{hyperref}

\newcommand{\ket}[1]{|\,#1\,\rangle}          % 
\newcommand{\ud}{{\mathrm{d}}}
\hfuzz=\maxdimen \tolerance=10000
\vfuzz=\maxdimen \tolerance=10000
\newcommand{\LCm}{{\scriptscriptstyle -}}
\newcommand{\LCp}{{\scriptscriptstyle +}}

\newcommand{\LCperp}{{\scriptscriptstyle \perp}}
\newcommand{\be}{\begin{equation}}
\newcommand{\ee}{\end{equation}}
\newcommand{\bi}{\begin{enumerate}}
\newcommand{\ei}{\end{enumerate}}

\newcommand{\sff}{{\mathsf{f}}}

\newcommand{\e}{{\mathrm{e}}}
\newcommand{\la}{{\langle}}
\newcommand{\ra}{{\rangle}}

\DeclareFontEncoding{LGR}{}{}
\DeclareSymbolFont{sfitgreek}{LGR}{cmss}{m}{it}
\DeclareMathSymbol{\sfpi}{\mathord}{sfitgreek}{`p}

\begin{document}

\title{Scattering and depletion in a flying focus from conformal transformations}

\author{Tim Adamo}
\email{t.adamo@ed.ac.uk}
\affiliation{School of Mathematics \& Maxwell Institute for Mathematical Sciences, University of Edinburgh, EH9 3FD, UK}

\author{Anton Ilderton}
\email{anton.ilderton@ed.ac.uk}
\affiliation{Higgs Centre, School of Physics and Astronomy, University of Edinburgh, EH9 3FD, UK}

\author{Adam Noble}
\email{adam.noble@strath.ac.uk}
\affiliation{Department of Physics, SUPA, University of Strathclyde, Glasgow, G4 0NG, UK}

\begin{abstract}
We show that flying focus fields can be obtained from complex conformal transformation of plane waves, and that solutions of the massless wave equation in the so-obtained fields are, correspondingly, conformal transformations of the Volkov solutions. This leads to the result that photon emission amplitudes in a totally depleting flying focus beam may be computed directly from the corresponding plane wave amplitudes by taking a simple Gaussian average over certain momentum variables. In effect, this gives a way of introducing focussing effects into strong-field QED calculations `for free'. The extension of these results to scattering amplitudes including only partial depletion is discussed and some first results presented in the anti-self-dual limit. 
\end{abstract}

\maketitle

\tableofcontents

\section{Introduction}

Modern lasers produce light of such intensity that they access the quantum and relativistic regime of light-matter interactions~\cite{Abramowicz:2021zja,Sarri:2025qng,Los:2024ysw}. Furthermore, the effective laser-matter coupling exceeds unity, demanding a non-perturbative treatment. The appropriate theory framework for laser-particle scattering is then `strong-field quantum electrodynamics (QED)', which employs background-field quantum field theory (QFT) methods to account for the strong laser coupling; for reviews see~\cite{Gonoskov:2021hwf,Fedotov:2022ely}.  

Performing analytic calculations in strong-field QED requires two main ingredients: solutions of Maxwell's equations which model the strong background field, and explicit electron wavefunctions which are exact in that background. Of course, there is the additional practical requirement that these objects (the background and its associated wavefunctions) are amenable to calculation -- that is, that scattering amplitudes and associated observables can actually be computed to reasonable levels of precision within the Furry expansion~\cite{Furry:1951bef}. This is one of the main bottlenecks to making analytic progress in strong-field QED, since there are few backgrounds which allow this.

For these reasons, plane waves provide one of the most ubiquitous examples of strong QED backgrounds, as the required charged particle wavefunctions (the Volkov solutions) are known for arbitrary plane waves as well as particle spin and charge~\cite{Wolkow:1935zz,Seipt:2017ckc}. However, intense electromagnetic fields created in the laboratory, or nature, typically feature spatial inhomogeneity, one example being the focussing geometry of ultra-intense lasers. Plane waves, though, are completely un-focussed, and will therefore miss physics resulting from focussing effects, see e.g.~\cite{Fedotov2009exact,Gies:2013yxa,Gonoskov:2013ada,Heinzl:2017zsr,Karbstein:2019dxo,PhysRevA.103.012215,DegliEsposti:2023qqu,Valialshchikov:2025gkj}. This has motivated interest in exact solutions of Maxwell's equations which go beyond the plane wave approximation by including focussing behaviour. Prominent examples are \emph{flying focus} fields, part of a wide class of solutions of Maxwell's equations with a focal spot that moves with or against the beam, see~\cite{Brittingham83,Sezginer1985,HILLION1988143,Hillion92} for some of the original investigations of such solutions, \cite{Froula2018,Jolly:20,FFexpt1,FFexpt2} for experiments, and~\cite{PhysRevA.103.012215,Formanek:2021bpw,Formanek:2023mkx,Adamo:2025vzv} for recent applications to strong-field QED processes.

However, it is not possible to obtain exact solutions for the wavefunctions of charged fields in general flying focus backgrounds, meaning that most studies have utilized numerics or perturbative approximations of the background. Nevertheless, it was recently shown that for certain \emph{complex-valued} flying focus solutions  it is possible to obtain exact, analytic expressions for massless scalar/ultra-relativistic electron wavefunctions~\cite{Adamo:2025vzv}. Furthermore, scattering amplitudes in these complex backgrounds have a \emph{real} physical interpretation, encoding the total depletion of an initial flying focus field (viewed as an incoming coherent state) during a scattering process~\cite{Endlich:2016jgc,Ilderton:2017xbj,Aoude:2023fdm}. We will here tie these ideas together, showing first that a flying focus beam can be obtained by applying a complex conformal transformation to a real plane wave. That the resulting field is complex is not an issue at this level -- one simply takes the real part to obtain the fields of a real flying focus.
Next, we will show that the same conformal transformation turns Volkov wavefunctions (exact solutions of the wave equation for charged particles in plane wave backgrounds~\cite{Wolkow:1935zz}) into the exact wavefunctions for charged particles in the flying focus background.

There are two caveats here -- first, this result holds only for massless charges, due to the transformation being conformal. Second, the operation of taking real parts and solving the Klein-Gordon (or Dirac) equations do not commute, meaning that we can only find wavefunctions in the \emph{complex} flying focus background.  This restricts us to considering amplitudes for processes in which the flying focus beam, again viewed as an incoming coherent state of photons, is completely absorbed during scattering. Despite this, we will be able to find \emph{all-multiplicity} results relating scattering in focussed fields to scattering in plane wave backgrounds.

While an interesting theory challenge, the case of complete depletion is not phenomenologically relevant for current experiments~\cite{Seipt:2016fyu}. This prompts us to ask how to account for only partial depletion, and we will provide some initial results in the latter part of this paper. Restricting now to \emph{anti-self-dual} flying focus beams, we find a set of exact charged particle wavefunctions which allow for arbitrary levels of depletion (staying within the anti-self-dual sector) and discuss their properties. Although only a first step, these results offer an interesting look into the depth of complexities one might expect to appear in wavefunctions for realistic, and real, focussed backgrounds. 

This paper is organised as follows. In Sec.~\ref{sec:review} we provide a short review of relevant results on scattering amplitudes involving coherent states, the relation of such amplitudes to scattering on backgrounds (real and complex), and essential properties of the Volkov wavefuntions describing charged particles scattering on plane wave backgrounds. In Sec.~\ref{sec:everythingconformal} we show that a flying focus beam can be constructed by taking a complex special conformal transformation of a plane wave, and that the same transformation turns Volkov solutions into wavefunctions for charged particles in flying focus backgrounds. We proceed to the computation of amplitudes in Sec.~\ref{sec:amplitudes}, beginning with the simplest case of nonlinear Compton scattering (with complete absorption). This will set the scene for our main all-multiplicity result: knowing the plane wave $N$-photon emission amplitude gives -- almost automatically -- the corresponding amplitude in a flying focus background. In Sec.~\ref{sec:partial-depletion} we begin the investigation of partial depletion amplitudes, restricting here to the case of anti-self-dual backgrounds, where we are able to determine an exact set of wavefunctions for which we expect an extra degree of simplicity. We conclude in Sec.~\ref{sec:conclusions}.

\subsubsection*{Notation and conventions}
We choose units such that $\hbar=c=1$. The metric is mostly minus. We absorb factors of $2\pi$ into measures and delta functions, writing $\hat\ud p:=\ud p/(2\pi)$ and $\hat\delta(p) := 2\pi \delta(p)$. The positive-frequency, 
on-shell phase space measure over massless particle momenta $\ell_\mu$ is denoted
\be\label{OPSmeasure}
\int_{\ell}\equiv
\int {\hat\ud}^4\ell \,\, \hat\delta(\ell^2)\,
\Theta(\ell_0)\;.
\ee
%%
%%%%%%%%%%%%%%%%%%%%%%%%%%%%%%%%%%%%%
%\clearpage
%%%%%%%%%%%%%%%%%%%%%%%%%%%%%%%%%%%%%
\section{Review: Coherent states, backgrounds and plane waves}\label{sec:review}
%%%%%%
In this section, we briefly review some familiar structures of strong-field QED in plane wave backgrounds, and recall that strong-field scattering amplitudes in complex-valued electromagnetic fields encode real depletion/backreaction effects in scattering between coherent states.

%%%%%%%%%%%%%%%%%%%%%%%%%%%%%%%%%%%%%

\subsection{Background fields and coherent states}
Our focus throughout will be on massless scalar QED, viewed as an ultra-relativistic approximation for QED proper. In the presence of a fixed, classical background $A$, the Lagrangian governing the Furry expansion of this theory is~\cite{Furry:1951bef}:
\be\label{BFLag}
\mathcal{L}=(D^\mu \phi)^{*} D_{\mu}\phi\,-i\,e\,\mathcal{A}_{\mu}\left(\phi^*\,D^{\mu}\phi-\phi (D^{\mu}\phi)^{*}\right)+e^2\,\mathcal{A}^2\,|\phi|^2-\frac{1}{4}\,\mathcal{F}_{\mu\nu}\,\mathcal{F}^{\mu\nu}\,,
\ee
in which $\phi$ is the complex, charged scalar, $D_{\mu}=\partial_\mu + i e A_\mu$ is the covariant derivative associated with the background, and $\mathcal{A}_{\mu}$ is the fluctuating photon field, with field strength $\mathcal{F}_{\mu\nu}=\partial_{\mu}\mathcal{A}_{\nu}-\partial_{\nu}\mathcal{A}_{\mu}$.

We denote the S-matrix of this theory by $\mathcal{S}[A]$, highlighting the dependence on the background. Then scattering amplitudes in the Furry picture correspond to matrix elements $\left\langle\mathrm{out}|S[A]|\mathrm{in}\right\rangle$, where the initial state $|\mathrm{in}\rangle$ and final state $|\mathrm{out}\rangle$ are composed of scalars and free photons. Now, suppose the background is a vacuum solution of Maxwell's equations which can be written in terms of on-shell photon modes as
\be\label{realCS}
A_{\mu}(x)=
\int_{\ell}\, \left[\varepsilon_{\mu}^{(h)}(\ell)\,\alpha_{h}(\ell)\,\e^{-i\,\ell\cdot x}+\varepsilon^{(-h)}_{\mu}(\ell)\,\alpha^{*}_{h}(\ell)\,\e^{i\,\ell\cdot x}\right]\,,
\ee
where $\varepsilon^{(h)}_{\mu}(\ell)$ are helicity state polarization vectors with $h=\pm1$ (repeated helicity labels are implicitly summed over), obeying $\ell\cdot\varepsilon^{(h)}(\ell)=0$ and $[\varepsilon^{(h)}_{\mu}(\ell)]^{*}=\varepsilon^{(-h)}_{\mu}(\ell)$; and $\alpha_{h}(\ell)$ are (as we will see) occupation numbers for photons of helicity $h$ and momentum $\ell$. 

It is a long-known result that scattering amplitudes calculated in the presence of a vacuum solution (\ref{realCS}) are closely and precisely related to scattering of \emph{coherent states}~\cite{Frantz:1965,Kibble:1965zza}, defined as follows.
Let $a^{\dagger}_h(\ell)$ denote the creation operator for a photon of helicity $h$ and momentum $\ell$, and $a_h(\ell)$ denote the corresponding annihilation operator. Then a coherent state $\ket{\alpha}$ is given by acting on the vacuum with the displacement operator $\mathbb{D}(\alpha)$, so $\ket{\alpha} = \mathbb{D}(\alpha)\ket{0}$ where
\be\label{DisplOp}
\mathbb{D}(\alpha)=\exp\!\left[\int_{\ell}\left(a_h^{\dagger}(\ell)\,\alpha_{h}(\ell)-a_h(\ell)\,\alpha^*_h(\ell)\right)  \right]\,,
\ee
and the $\alpha_h(\ell)$ are, again, two freely chosen functions of momentum and helicity. The choice of notation here is deliberate, because in terms of the classical field \eqref{realCS}, it is a remarkable fact that the displacement operator can be used to translate the vacuum S-matrix, $S$ (i.e., with no background field) to the strong-field S-matrix $S[A]$ via:
\be\label{Flat2Strong}
S[A]=\mathbb{D}^{\dagger}(\alpha)\,S\,\mathbb{D}(\alpha)\,.
\ee
In terms of scattering amplitudes, this has the consequence that
\be\label{Flat2StrongAmps}
\left\langle\mathrm{out}\,|\,S[A]\,|\,\mathrm{in}\right\rangle
=
\left\langle\mathrm{out}\,|\,\mathbb{D}^{\dagger}(\alpha)\,S\,\mathbb{D}(\alpha)\,|\,\mathrm{in}\right\rangle=\left\langle\mathrm{out\,;}\,\alpha\,|\,S\,|\mathrm{in\,;}\,\alpha\right\rangle\,.
\ee
In other words, the background field amplitude $\left\langle\mathrm{out}\,|\,S[A]\,|\,\mathrm{in}\right\rangle$ is equal to the vacuum probability amplitude to evolve from an incoming state $|\mathrm{in}\rangle$ to an outgoing state $|\mathrm{out}\rangle$ in the presence of a given coherent state of photons, $|\alpha\rangle$.

%%%%%%%%%%%%%%%%%%%%%%%%%%%%%%%%%

\subsection{Complex backgrounds and depletion amplitudes}\label{sec:complexbackgrounds}
One expects that scattering off an initial coherent state should result, in realistic settings, in a \emph{different} final coherent state, due to e.g.~depletion or backreaction effects. Scattering amplitudes for such processes are given by
\be\label{CStrans1}
\left\langle\mathrm{out\,;}\,\beta\,|\,S\,|\mathrm{in\,;}\,\alpha\right\rangle\,,
\ee
with $|\alpha\rangle$ the incoming coherent state and $|\beta\rangle$ the outgoing coherent state. (Independent of the physical interpretation, these are in any case sensible amplitudes to consider as, for example, coherent states provide an overcomplete basis of all states.)  Using basic properties of the displacement operator \eqref{DisplOp} and restricting attention to incoming/outgoing number states which have zero kinematic overlap with the outgoing/incoming coherent states\footnote{See~\cite{Ilderton:2017xbj} for extensions.}, it follows that 
\be\label{CStrans2}
\left\langle\mathrm{out\,;}\,\beta\,|\,S\,|\mathrm{in\,;}\,\alpha\right\rangle= \exp\!\left[-\frac{1}{2}\int_{\ell}\left(|\alpha_h(\ell)|^2+|\beta_h(\ell)|^2-2\,\alpha_{h}(\ell)\,\beta_{h}^{*}(\ell)\right)\right] \left\langle\mathrm{out}|\,S[A]\,|\mathrm{in}\right\rangle\,,
\ee
where the classical potential $A_{\mu}$ defining the background field S-matrix $S[A]$ is now \emph{complex}:
\be\label{complexCS}
A_{\mu}(x)=
\int_{\ell}\, \left[\varepsilon_{\mu}^{(h)}(\ell)\,\alpha_{h}(\ell)\,\e^{-i\,\ell\cdot x}+\varepsilon^{(-h)}_{\mu}(\ell)\,\beta^{*}_{h}(\ell)\,\e^{i\,\ell\cdot x}\right]\,.
\ee
In other words, amplitudes with different coherent pieces in the initial and final states are equivalent (up to an overall exponential factor which is \emph{not} a phase) to amplitudes in a complex-valued background~\eqref{complexCS}, whose positive and negative frequency modes are given by the incoming and outgoing coherent state profiles, respectively. This observation provides a theoretical framework to compute depletion amplitudes (i.e.~scattering between distinct asymptotic coherent states) in terms of background field amplitudes where the background is complex. For example, a \emph{total depletion} process, whereby the incoming coherent state is totally absorbed during the scattering process, is modelled by a complex background~\eqref{complexCS} with $\alpha_h(\ell)\neq 0$, $\beta_{h}(\ell)=0$. 

Note that despite the appearance of a complex background in the Furry picture, the initial transition amplitude between coherent states is defined with respect to the ordinary S-matrix of (scalar) QED, so unitarity is preserved.

%%%%%%%%%%%%%%%%%%%%%%%%%%%%%%%%%%%%%

\subsection{Plane waves and Volkov wavefunctions}
It is illustrative and useful to consider the example of plane wave backgrounds, and the Volkov solutions of the Klein-Gordon equation in them.
Let the Minkowski line element in lightfront coordinates $(x^\LCp,x^\LCm,x^1,x^2)$ be
\be\label{Minkmet}
\ud s^2 = 2\, \ud x^\LCp \ud x^\LCm - |\ud x^\LCperp|^2 \;, \quad \qquad x^{\LCperp}=\left(x^1,\,x^2\right) \;,
\ee
with $|x^\LCperp|^2\equiv (x^1)^2+(x^2)^2$. Then plane wave Maxwell fields may be described by gauge potentials
\be\label{PlaneWavepot}
A_{\text{pw}}(x)=-x^\LCperp\, {\dot f}_{\LCperp}(x^\LCm)\, \ud x^\LCm\;, \qquad
\dot{f}(x^\LCm)\equiv\frac{\ud f(x^\LCm)}{\ud x^\LCm} \,,
\ee
where $\dot{f}_{\LCperp}(x^\LCm)$ are two freely chosen functions of lightfront time $x^\LCm$. Minus signs and derivative are for convenience. 
It is easy to see that \eqref{PlaneWavepot} solves the vacuum Maxwell equations for any choice of $f_{\LCperp}(x^\LCm)$, \emph{real or complex}, though  physical considerations can be used to place some restrictions on their form. For instance, demanding that $\dot{f}_{\LCperp}$ be compactly supported in $x^\LCm$ gives rise to `sandwich' plane waves which have well-defined asymptotic in- and out- regions for charged particle scattering, and for which the potential conveniently vanishes asymptotically~\cite{Schwinger:1951nm}.

Furthermore, it is a fact that \emph{any} electromagnetic field in Minkowski spacetime is well-approximated by a plane wave in the neighbourhood of a null geodesic~\cite{Penrose:1976}, and the high degree of symmetry preserved by plane waves -- a five-dimensional Heisenberg algebra whose centre is generated by translations in the $x^\LCp$-direction~\cite{Bergmann:1978fi,Trautman:1980bj,Heinzl:2017blq,Heinzl:2017zsr,Adamo:2017nia} -- means that wave equations for charged fields on plane wave backgrounds are \emph{exactly} solvable. For the simple example of a massless charged scalar $\phi(x)$, viewed as an approximation to an ultra-relativistic electron, the appropriate equation is Klein-Gordon:
\be\label{cKGeq}
D^2_{\text{pw}}\phi \equiv\left(\partial+i\,e\,A_{\text{pw}}\right)^2\phi=0\;.
\ee
The solutions of this equation are the well-known Volkov wavefunctions~\cite{Wolkow:1935zz} given by, writing $f_\mu \equiv \delta_\mu^\LCperp f_\LCperp$,
\be\label{Volkov}
\phi(x)=\e^{-i\,S_p(x)}\,, \qquad S_p(x)=- ef(x^\LCm)\cdot x +p\cdot x  + \frac{1}{2\,p_+}\int^{x^\LCm}_{-\infty}\!\!\ud s\left(2\,e f(s)\cdot p-e^2 f^2(s)\right)\,,
\ee
where $p_{\mu}$ is an on-shell 4-momentum obeying $p^2=0$. Similar solutions exist for massive fields with arbitrary spin~\cite{Seipt:2017ckc}.

The exact Volkov solutions and the symmetry properties of the plane wave background mean that scattering amplitudes computed in the Furry expansion can be obtained fairly explicitly, even for generic wave profiles $f_{\LCperp}(x^\LCm)$, although for every vertex there is an integral over lightfront time $x^\LCm$ which typically cannot be performed exactly. (See e.g.~\cite{Gonoskov:2021hwf,Fedotov:2022ely} for reviews of various approximation schemes.)  Notable exceptions are constant fields~\cite{Ritus:1985vta}, and the impulsive limit, for which closed-form, exact expressions for scattering probabilities have been obtained for non-linear Compton and pair creation at tree-level~\cite{Fedotov:2013uja,Ilderton:2019vot}, as well as spin/helicity-flip at one loop~\cite{Ilderton:2024ufp,AlexanderNew}.

%%%%%%%%%%%%%%%%%%%%%%%%%%%%%%%%%%%%%
%%%%%%%%%%%%%%%%%%%%%%%%%%%%%%%%%%%%%

\section{Focussed fields \& wavefunctions from conformal transformations}\label{sec:everythingconformal}
In this section, we show that the origin of the flying focus field and its associated wavefunctions can be understood as a complex-valued conformal transformation of familiar structures in plane wave backgrounds. Of course, the method of generating new solutions to massless wave and Maxwell equations via conformal transformations goes back to Bateman~\cite{Bateman:1909pyp}, and the observation that flying focus fields can be generated from complex conformal transformations has also been made (using somewhat different terminology) previously~\cite{Hillion92}. However, the fact that these conformal transformations can also be used to generate exact solutions for massless wavefunctions on flying focus backgrounds corresponding to total depletion processes is new. (For applications of conformal transformations to radiation reaction see e.g.~\cite{Hobbs,Noble:2021ebw}.)

%%%%%%%%%%%%%%%%%%%%%%%%%%%%%%%%%%%%

\subsection{Flying focus from a conformal transformation}\label{sec:build-FF}

Conformal transformations of Minkowski spacetime rescale the metric \eqref{Minkmet} by an overall positive factor, $\ud s^2\to\Omega^2(x)\,\ud s^2$, which preserves angles but not scales. These transformations form the group SO$(2,4)$, which is generated by Poincar\'e translations, Lorentz transformations, dilatations (or scale transformations) and the special conformal transformations, which act as
\be\label{SCTrans}
x^{\mu}\to \frac{x^{\mu}-b^{\mu}\,x^2}{1-2\,b\cdot x+b^2\,x^2}\,,
\ee
for some 4-vector $b^{\mu}$. When considering physical theories which are conformally invariant, conformal transformations can be used to generate interesting new solutions to equations of motion from known ones. Now, the vacuum Maxwell equations are famously conformally invariant~\cite{Cunningham:1910pxu,Bateman:1910mvi}, so the transformation \eqref{SCTrans} will send, for example, a plane wave Maxwell field to a \emph{new} solution of the vacuum Maxwell equations, with spatial inhomogeneity.

Consider the specific case of a special conformal transformation \eqref{SCTrans} in which $b^{\mu}$ is chosen to be a null vector such that $2b\cdot x=\kappa\,x^+$ with $\kappa\in\mathbb{R}$. This special conformal transformation defines a new set of coordinates
\be\label{realConfTrans}
x^{\mu}\rightarrow\sigma^{\mu}\,, \qquad \sigma^{\LCm}=x^\LCm+\frac{\kappa}{2}\,\frac{|x^{\LCperp}|^2}{1-\kappa\,x^\LCp}\,, \quad \sigma^{\LCperp}=\frac{x^{\LCperp}}{1-\kappa x^+}\,, \quad \sigma^\LCp=\frac{x^\LCp}{1-\kappa\,x^+}\,.
\ee
Suppose we apply this special conformal transformation to a plane wave. Before giving any details, let us first consider some broad features of the new Maxwell field. The solution will have the attractive feature of introducing spatial inhomogeneity: whereas the initial plane wave was totally homogeneous in the transverse plane (being a function of lightfront time $x^\LCm$ only), the new solution depends on the $x^\perp$-coordinates through the combination $\sigma^-$ given in \eqref{realConfTrans}. Unfortunately, the resulting field will almost inevitably be singular, blowing up along the lightfront where $x^\LCp=\kappa^{-1}$. This can be interpreted as a sort of focussing caustic induced by the conformal transformation.

This singular behaviour can be avoided if we allow ourselves to consider \emph{complex} conformal transformations. Indeed, replacing $\kappa$ in \eqref{realConfTrans} with $-ik$ for some real number $k\geq 0$ leads to 
\be\label{compConfTrans}
x^{\mu}\rightarrow\sigma^{\mu}\,, \qquad \sigma^{\LCm}=x^\LCm -\frac{i\,k}{2}\,\frac{|x^{\LCperp}|^2}{1+i\,k\,x^\LCp}\,, \quad \sigma^{\LCperp}=\frac{x^{\LCperp}}{1+i\,k x^\LCp}\,, \quad \sigma^\LCp=\frac{x^\LCp}{1+i\,k\,x^\LCp}\;.
\ee
The apparent downside of such a choice is that the conformal transformation of a real Maxwell field will inevitably become complex. However, we have already seen in Sec.~\ref{sec:complexbackgrounds} how to interpret such fields, and so we will now apply (\ref{compConfTrans}) to the plane wave (\ref{PlaneWavepot}) and explore the newly generated field. The special conformal transformation of (\ref{PlaneWavepot}) is
\be\label{PlaneWavePotTranformed}
\begin{split}
    A(x) &=-\sigma^\LCperp\, {\dot f}_{\LCperp}(\sigma^\LCm)\, \ud \sigma^\LCm\ \\
    &= \frac{-x^\LCperp {\dot f}_\LCperp(\sigma^\LCm)}{1+i\, k\, x^\LCp} \bigg(\ud x^\LCm  - \frac{i\,k\, x^\LCperp\, \ud x^\LCperp}{1+i\, k\, x^\LCp}  - \frac{k^2}{2} \frac{|x^\LCperp|^2}{(1+i\, k\, x^\LCp)^2}\,\ud x^\LCp \bigg)\;. 
\end{split}
\ee
This is easily confirmed to be a vacuum solution of Maxwell's equations, but we need to do some work to clarify its physical behaviour. Let us consider first the functions $\dot{f}_\LCperp(\sigma^\LCm)$, which we write in terms of their Fourier modes as
\be
    {\dot f}_\LCperp(\sigma^\LCm) = \int_0^\infty\!\frac{\ud \omega}{2\pi} (-i\omega) f_\LCperp(\omega)\, \e^{-i\omega \big[x^\LCm - \frac{ik}{2}\frac{|x^\perp|^2}{1+i k x^+}\big]} + \int_0^{\infty}\!\frac{\ud \omega}{2\pi} (+i\omega) f_\LCperp(-\omega)\, \e^{i\omega  \big[x^\LCm - \frac{ik}{2}\frac{|x^\perp|^2}{1+i k x^+}\big]} \;.
\ee
We have deliberately separated the field into positive energy/incoming and negative energy/outgoing modes to highlight the following: the exponent is not a pure phase, there is a real part equal to
\be
    \mp\frac{\omega k}{2} \frac{|x^\LCperp|^2}{1+k^2\, (x^\LCp)^2} \;,
\ee
representing either an exponential damping or exponential increase. Thus, the Maxwell field (\ref{PlaneWavePotTranformed}) resulting from the conformal transformation~\eqref{compConfTrans} is only physically sensible (i.e., the Fourier integrals do not diverge due to exponentially increasing contributions) for purely positive-frequency, i.e.~incoming, profiles $f_\LCperp$.

We therefore restrict to this case. With this we have a sensibly defined, albeit still complex, background. From Sec.~\ref{sec:complexbackgrounds} we know, though, that this is equivalent to the presence of an \emph{initial} coherent state, and therefore an initially present (real!) field in our scattering process. The associated field is just the real part of (\ref{PlaneWavePotTranformed}).  To establish the corresponding coherent state profile, we first observe that the transformed gauge potential may be written
\be\label{ComplexFFgauged}
    A=-\ud\big(\sigma^\LCperp f_\LCperp(\sigma^\LCm)\big) + \frac{f_\LCperp(\sigma^\LCm)}{1+i\,k\,x^+}\bigg(\ud x^\LCperp-\frac{i\,k\,  x^\LCperp}{1+i\,k\,x^+}\,\ud x^\LCp\bigg)\,,
\ee
in which the first term is pure gauge. The second term may easily be verified to admit a Fourier transform of the form (\ref{complexCS}) with ($\beta=0$ and)
\be\label{CpxFF-CohState}
    \alpha_h(\ell)=\frac{4\pi}{k}\,\exp\!\left(-\frac{|\ell_{\LCperp}|^2}{2\,k\,\ell_\LCm}\right)\,f_{h}(\ell_{\LCm})\,,
\ee
in which $f_{h} := (f_1 -i\,h\, f_2)/\sqrt{2}$, and where we work in (anti-)lightfront gauge $\varepsilon^{(h)}_{-}=0$, in which case the polarizations have the explicit form\footnote{To be clear, inserting (\ref{photonpol}) and (\ref{CpxFF-CohState}) into (\ref{complexCS}) and performing the Fourier transform recovers the second, nontrivial term in (\ref{ComplexFFgauged}). The first term in (\ref{ComplexFFgauged}) represents a large gauge transformation; it can, along with the equivalent potential (\ref{PlaneWavePotTranformed}), be accommodated in the coherent state picture but this requires more complicated polarisations -- see~\cite{Cristofoli:2022phh} for a recent discussion.}
\be\label{photonpol}
    \varepsilon_{\mu}^{(h)}(\ell)
   = \frac{1}{\sqrt{2}}\left(\frac{\ell_1 +i\, h\, \ell_2}{\ell_\LCm},0,1, i h\right)\;.
\ee
The coherent state profile (\ref{CpxFF-CohState}) is easily understood; the Gaussian momentum distribution has variance proportional to $k$, which in the limit $k\to 0$ recovers a transverse delta function in $\ell_\LCperp$, which is the Fourier profile (occupation number distribution) of a plane wave. In other words, our conformally transformed field is obtained from a plane wave by broadening the spread of transverse photon momenta -- this is, as previously observed~\cite{Sezginer1985}, precisely how one constructs a \emph{flying focus} beam from a plane wave (see also~\cite{DiPiazza:2024wxt,Ramsey:2026yil} for applications). 

To be concrete, let us take
\be\label{FFBeam}
    f_\LCperp(x^\LCm) = \epsilon_\LCperp \e^{-i\omega x^\LCm}
\ee
in which $\epsilon_\LCperp$ is a (possibly complex) constant vector, while $\omega$ is a fixed positive frequency. The critical factor appearing in the resulting $A_\mu$ and $F_{\mu\nu}$ is
\be\label{FF-beam-choice}
    \frac{\e^{-i\omega \sigma^\LCm(x)}}{1+i\,k\,x^\LCp} \;,
\ee
and it is enough to analyse this. Taking the real part of this factor we find 
\be\label{RePhi0}
\frac{\e^{-\frac{\omega k}{2}\,\frac{|x^\LCperp|^2}{(1+k^2\,(x^\LCp)^2)}}}{1+k^2\,(x^\LCp)^2}
\left[\cos\!\left(\omega\left[x^\LCm -\frac{k^2\,x^\LCp\,|x^\LCperp|^2}{2\,(1+k^2\,(x^\LCp)^2)}\right]\right)-k\,x^\LCp\,\sin\!\left(\omega\left[x^\LCm-\frac{k^2\,x^\LCp\,|x^\LCperp|^2}{2\,(1+k^2\,(x^\LCp)^2)}\right]\right)\right]\,.
\ee
The gauge field thus describes a Gaussian peak counter-propagating against a wave of frequency $\omega$ at the speed of light. (In fact (\ref{RePhi0}) is itself a solution of the scalar wave equation.)

\begin{figure}[t!]
    \centering
    \includegraphics[width=0.3\linewidth]{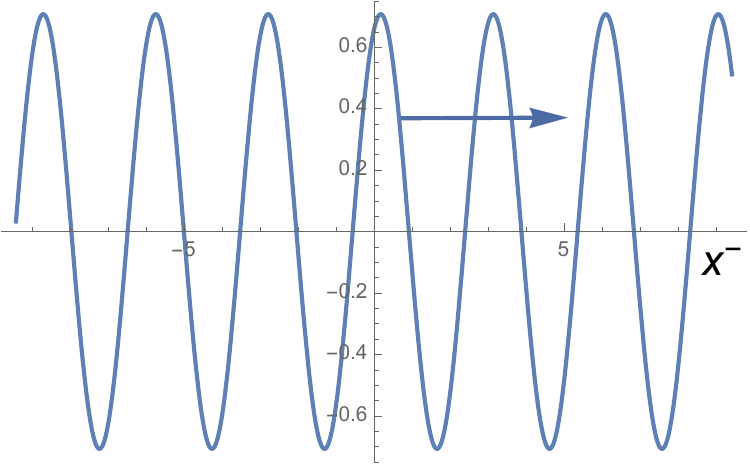}
    \quad
    \raisebox{5pt}{\includegraphics[width=0.3\linewidth]{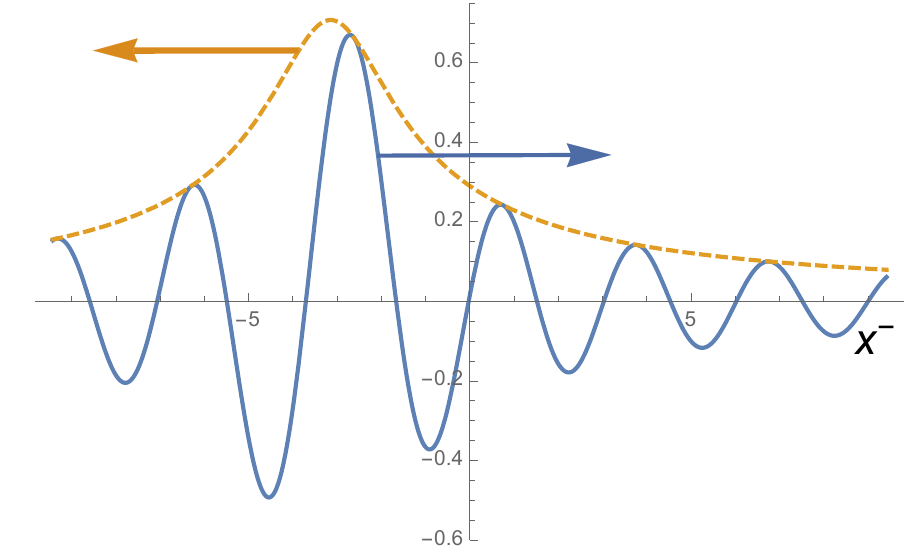}}
    \quad
    \raisebox{-15pt}{\includegraphics[width=0.3\linewidth]{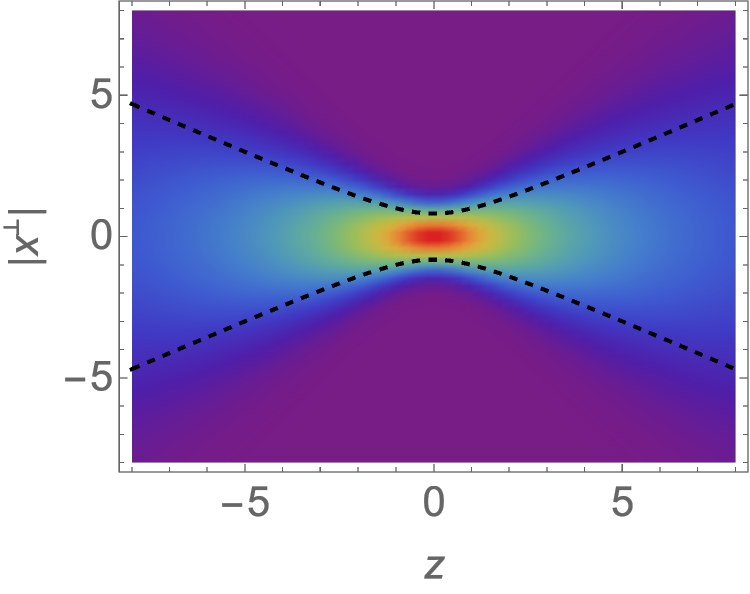}}
    \caption{\label{Fig:FF-plots}
Left: sketch of the `seed' plane wave electric fields corresponding to (\ref{FFBeam}), plotted as a function of $x^\LCm$, equivalently at fixed $t$ and as a function of $z$. Middle: electric field component of the flying focus beam at $x^\LCperp=0$ -- a Gaussian peak (illustrated by the dashed yellow line) now counter-propagates with the wave. Right: density plot of the electric field strength $|{\bf E}|$ in the flying focus beam, at $t=0$, as a function of $z$ and $|x^\LCperp|$. The spot size $w(x^\LCp)$ from (\ref{w-and-R}) is marked by the dashed line.}
\end{figure}

In analogy to the case of Gaussian beams\footnote{We remind the reader that standard Gaussian beam expressions are approximate (paraxial) solutions of Maxwell's equations, whereas the solutions here are exact.} we can read off the waist radius of the Gaussian peak $w_0$ and its Rayleigh length $z_R$ in terms of $\omega$ and $k$ as
\be
    w_0 = \sqrt{\frac{2}{\omega k}} \;, \qquad z_R = \frac{1}{k} \;.
\ee
The spot size $w(x^\LCp)$ and curvature $R(x^\LCp)$ are given by
\be\label{w-and-R}
w(x^\LCp) = w_0 \sqrt{1+ \left(\frac{x^\LCp}{z_R}\right)^2} \;,
\qquad
    R(x^\LCp) = x^\LCp \left(1 + \left(\frac{z_R}{x^\LCp}\right)^2\right)\,,
\ee
which again are the standard expressions for a Gaussian beam, except they now depend on $x^\LCp$ rather than $x^3$.  The fields are illustrated in Fig.~\ref{Fig:FF-plots}.

This solution of Maxwell's equations, which follows from the choice (\ref{FF-beam-choice}) of a monochromatic profile, is called the \emph{flying focus beam} solution of Maxwell's equations. It is the simplest example in a broad class of flying focus solutions -- for other examples with e.g.~angular momentum and different focal velocities see~\cite{Ramsey2023,Formanek:2023zia}. 
For choices of profile function $f_\LCperp$ other than (\ref{RePhi0}) the resulting fields have been referred to as `focus waves'~\cite{Sezginer1985}. We will, for simplicity, continue to refer to the general solution  (\ref{PlaneWavePotTranformed}) obtained by conformal transformation as a flying focus solution, although the physical properties of the field naturally depend on the choice of $f_\LCperp(x^\LCm)$.
%

%%%%%%%%%%%%%%%%%%%%%%%%%%%%%%%%%%%

\subsection{Massless wavefunctions from a conformal transformation}\label{sec:phi-from-conformal}

The complex flying focus solutions above present an opportunity to analytically compute scattering amplitudes which include the effects of both depletion \emph{and} spatial inhomogeneity. However, to do so requires solutions of the charged wave equations in this complex background. In generic flying focus Maxwell fields, there are no known exact solutions, however it is possible to write down \emph{exact} wavefunctions in the flying focus~\eqref{PlaneWavePotTranformed}, by again using the complex conformal transformation \eqref{compConfTrans}, as follows. We again consider a massless, complex charged scalar $\Phi$, viewed as an ultrarelativistic description of electrons/positrons. If we act with \eqref{compConfTrans} then the Klein-Gordon equation on a plane wave background becomes, up to a conformal factor, the wave equation on a complex flying focus background, that is
\be\label{D2-transform}
    D^2(x) = \Omega^{-3} D^2_{\text{pw}}(\sigma) \Omega\;,
\ee
in which $D_{\text{pw}}^\mu$ is the plane-wave covariant derivative and the `conformal factor' $\Omega$ is defined by $\Omega(x)=1+i\, k\, x^\LCp$, equivalently $\Omega(\sigma) = 1/(1-i\,k\,\sigma^\LCp)$. 

It follows that if $\Phi$ is a solution of the Klein-Gordon equation in a plane wave background, then after the conformal transformation (\ref{compConfTrans}) $\hat{\Phi} \equiv \Omega^{-1}\Phi$ is a solution of the wave equation in the complex flying focus gauge field. (In other words the scalar has conformal weight -1.) Even more simply, the latter is solved by simply taking the conformal transformation of the Volkov solutions (\ref{Volkov});
\be
    D^2(x)\, \Omega^{-1}\Phi(\sigma(x)) = \Omega^{-3} D^2_{\text{pw}}(\sigma) \Phi(\sigma) = 0\;.
\ee
We can thus write down the solutions of the wave equation in the complex flying focus background (\ref{PlaneWavePotTranformed}) as
\be\label{ComplexFF-Volkov}
\begin{split}
    \hat{\Phi}_p(x) &= \Omega^{-1}(x)\, \e^{-i\, S_p(\sigma(x))} \\
    &=
    \frac{1}{1+i\, k\, x^\LCp} \exp\bigg[i ef(\sigma^\LCm)\cdot \sigma -ip\cdot \sigma  -\frac{i}{2\,p_+}\int^{\sigma^\LCm}\!\!\!\ud s\left(2\,e f(s)\cdot p-e^2 f^2(s)\right)\bigg]\bigg|_{\sigma\equiv \sigma(x)} \\
    &= 
    \frac{1}{1+i\, k\, x^\LCp} \exp\bigg[ i\frac{ef(\sigma^\LCm(x))\cdot x}{1+i\,k\,x^\LCp} -i \frac{p_\LCp x^\LCp + p_\LCperp x^\LCperp}{1+i\,k\,x^\LCp} - ip_\LCm\sigma^\LCm(x)  -\frac{i}{2\,p_+}\int^{\sigma^\LCm(x)}\!\!\!\ud s\left(2\,e f(s)\cdot p-e^2 f^2(s)\right)\bigg] \;,
\end{split}
\ee
in which we have not written out $\sigma^\LCm(x)$ explicitly for brevity.

These solutions preserve several features from the plane-wave case.
For example, despite clearly depending nontrivially on all spacetime coordinates, they are still semiclassical/WKB exact, in that the exponent is exactly of order $1/\hbar$. (The conformal prefactor essentially takes the place of an $\hbar^0$ term.) The exponent continues to obey
\be\label{Ham_Jac}
(\partial S_{p}-e\,A)^2=0\,,
\ee
that is, it solves the charged Hamilton-Jacobi equations on the complex flying focus background.

Let us comment on boundary conditions. For pulsed plane wave fields we may always choose $f(x^\LCm)\to0$ as $x^\LCm\to -\infty$, in which case (\ref{Volkov}) has incoming boundary conditions -- it describes a free electron of momentum $p_\mu$ in the asymptotic past. The wavefunction for an outgoing electron, momentum $q$ in the future, is obtained by replacing $p\to q$, setting the integral limit to $+\infty$, and replacing~\cite{Kibble:1965zza,Dinu:2012tj}
\be\label{memory}
    f(x^\LCm)\to f_\text{out}(x^\LCm):=f(x^\LCm)-f_\infty\,, \qquad f_{\infty}:=\lim_{x^-\to\infty}f(x^-) \;.
\ee
Finally, we take the conjugate, as this is how the wavefunction appears in $S$-matrix elements. Performing the conformal transformation one obtains the corresponding outgoing flying focus wavefunction (dropping the hat -- from hereon all our wavefunctions are those in flying focus backgrounds)
\be\label{ComplexFF-VolkovConj}
\begin{split}
    {\bar\Phi}_q(x) &=
    \frac{1}{1+i\, k\, x^\LCp} \exp\bigg[-i ef_{\text{out}}(\sigma^\LCm)\cdot \sigma +iq\cdot \sigma  +\frac{i}{2\,q_+}\int^{\sigma^\LCm}_\infty\!\!\!\ud s\left(2\,e f_{\text{out}}(s)\cdot q-e^2 f_{\text{out}}^2(s)\right)\bigg]\bigg|_{\sigma\equiv\sigma(x)} \;.
\end{split}
\ee
One might hope to obtain similar Volkov-like solutions for massive scalars and Dirac spinors. However, the equations of motion obeyed by these fields are not conformally invariant, due to the scale set by their masses. As such, it is not possible to generate exact solutions for their wavefunctions using the same conformal transformation. Nevertheless, we are now armed with the exact, ultrarelativistic wavefunctions \eqref{ComplexFF-Volkov} on an incoming flying focus field, and can thus turn to the computation of total depletion amplitudes in these backgrounds.

%%%%%%%%%%%%%%%%%%%%%%%%%%
%%%%%%%%%%%%%%%%%%%%%%%%%%

\section{Total depletion amplitudes in focussed fields}\label{sec:amplitudes}
We have now seen that applying a conformal transformation to a plane wave electromagnetic field turns it into a flying focus, and that the same transformation turns Volkov solutions into solutions of the Klein-Gordon equation in the flying focus background. The next question to address is what happens if we apply the same transformation to scattering amplitudes in plane wave backgrounds. We will do this in Sec.~\ref{sec:amplitude-transform}. As a warm-up, we will first consider nonlinear Compton scattering, and calculate its amplitude following the method of~\cite{Adamo:2025vzv}.

\subsection{Warm up: Nonlinear Compton scattering}

We consider the tree-level scattering of an electron on the flying focus background. The electron scatters from momentum $p$ to momentum $q$, and emits a photon of momentum $\ell$ and polarisation $\varepsilon^{(h)}_\mu(\ell)$. This is nonlinear Compton (NLC) scattering with total depletion.

To calculate the amplitude, let $\Phi_p$ and ${\bar \Phi}_q$ be, respectively, incoming and outgoing wavefunctions (\ref{ComplexFF-Volkov}) and (\ref{ComplexFF-VolkovConj}) for electrons of momentum $p$ and $q$.  It is easily checked that the current
\be\label{eq-j}
    j_\mu(x) = {\bar \Phi}_q D_\mu \Phi_p - \Phi_p {\bar D}_\mu{\bar \Phi}_q \;,
\ee
with ${\bar D}_\mu:= \partial_\mu - i\,e A_\mu$, is conserved as expected, $\partial_\mu j^\mu(x) =0$. In terms of this current, the $S$-matrix element for NLC scattering is
\be\label{NLC-amp}
    S_{fi} =  
    e\,\e^{W[\alpha]}
    \int\!\ud^4 x\, 
    \mathrm{e}^{i\ell\cdot x}\,\varepsilon_\mu^{(-h)}(\ell)\,
    j^\mu(x) \;,
\ee
which takes the same form as in a real background, aside from the factor $\e^{W[\alpha]}$, which is shorthand for the exponential in~(\ref{CStrans2}). We will perform the integrals over $x^\mu$ in (\ref{NLC-amp}) by deforming them in the complex plane, following~\cite{Adamo:2025vzv}; an alternative method will also be described below. We first displace the $x^\LCm$ contour from the real axis by the non-trivial part of $\sigma^\LCm$,
\be
    x^\LCm \to x^\LCm + \frac{ik}{2}\frac{|x^\LCperp|^2}{1+i k x^\LCp}\;,
\ee
and integrate along a real slice of $x^\LCm$. This change of variable causes a shift in the exponent of the photon wavefunction, and as a result the $x^\LCperp$--integrals become Gaussian. We rotate the $x^\LCperp$ contours in the complex plane, effectively changing variable to $\sigma^\LCperp=x^\LCperp/(1+i k x^\LCp)$. Performing the integrals generates the factor $\Delta_\LCperp(q+\ell+e f_\infty-p)$ where
\be\label{goodGauss}
        \Delta_\LCperp(p) := \frac{(2\pi)^2}{2k\ell_\LCm} \exp\bigg[-\frac{|p_\LCperp|^2}{2k \ell_\LCm}\bigg]
\ee
is a $k$-regulated delta function in the transverse (momentum) plane obeying
\be
        \lim_{k\to 0} \Delta_\LCperp(p) = {\hat \delta}^2(p_\LCperp)\;.
\ee
This should be contrasted with the plane wave calculation -- the exponent appearing in the integrand of the amplitude is there linear in $x^\LCperp$, which generates an exact delta function conserving overall transverse momenta. 

The next step is to perform the $x^\LCp$ integral. Changing variable once more from $x^\LCp$ to $\sigma^\LCp=x^\LCp/(1+i k x^\LCp)$, and integrating along real $\sigma^\LCp$, we obtain a delta function setting 
    \be\label{def:qstar}
        q_\LCp = q_\LCp^\star := p_\LCp - \frac{|p_\LCperp-q_\LCperp|^2}{2 \ell_\LCm} \;.
    \ee
(We pause here to comment that the contour deformations used above are valid provided they do not cross any singularities of the integrand, in particular of the background profile~$f(\sigma)$. This is certainly the case for the flying focus beam~(\ref{FFBeam}), see also~\cite{Adamo:2025vzv}. For more general $f(\sigma)$, we observe that no obvious pathologies have been encountered -- if, for example, the sign of the exponent in~(\ref{goodGauss}) had been positive or ambiguous, it would be a clear warning sign, but this is not the case. Similarly, the change of variable from $x^\LCp$ to $\sigma^\LCp$ leads to a real phase and therefore a delta function, and not an unphysical positive exponential. Below, we will give an alternative method of calculating amplitudes which does not rely on contour deformation arguments.)

Only the integral over $\sigma^\LCm$ remains, and cannot be evaluated in general, just as for the integral over $x^\LCm$ in a general plane wave background. We can however simplify by dropping any term which is a total derivative with respect to $\sigma^\LCm$~\cite{Boca:2009zz,Dinu:2012tj}, which allows us to tidy the final expression somewhat. The amplitude is conveniently expressed in terms of the `dressed' on-shell momenta
\be
    \begin{split}
        \pi^\mu_p(s) &:= p^\mu - e f^\mu(s) + n^\mu \frac{2ef(s)\cdot p-e^2f(s)^2}{2p_\LCp} \;,
        \\
        \pi^\mu_q(s) &:= q^\mu - e f_\text{out}^\mu(s) + n^\mu \frac{2ef_\text{out}(s)\cdot q-e^2f_\text{out}(s)^2}{2q_\LCp} \;,
    \end{split}
\ee
which are functionally identical to those appearing in plane wave backgrounds. They inherit the boundary conditions of the corresponding in and out modes, that is $\pi_p(-\infty)=p$ and $\pi_q(+\infty) = q$. We will write, for the transverse components of these momenta, $\sfpi(s) \equiv (\pi_1(s) + i \pi_2(s))/\sqrt{2}$; these are just the projections onto the transverse photon polarisation vectors, see (\ref{photonpol}). 

Finally then, the amplitude is
\be\label{NLC-result}
\begin{split}
   S_{fi} = 2ie\, 
   \ell_\LCm\,
   &{\hat\delta}(q_\LCp-q_\LCp^\star)\,\e^{W[\alpha]}\,
   {\Delta}_\LCperp(\ell+q +e f_\infty -p)\\
   &\int\!\ud \sigma\,
   \bigg[
    \delta_{h}^{+1}
        \frac{q_\LCp \sfpi_p(\sigma) - p_\LCp \sfpi_q(\sigma)}{(\sfpi_q(\sigma) - \sfpi_p(\sigma))^2}
    +\delta_{h}^{-1}
        \frac{q_\LCp {\bar\sfpi_p(\sigma)} - p_\LCp {\bar\sfpi}_q(\sigma)}{({\bar\sfpi}_q(\sigma) - {\bar\sfpi}_p(\sigma))^2}
    \bigg]
   \exp\bigg(
   \frac{i}{2}\int^{\sigma}\!\!\!\ud s\, \frac{(\pi_q(s)-\pi_p(s))^2}{q_\LCp-p_\LCp}\bigg) \;.
\end{split}
\ee
In the second line, we have written the pre-exponential factor as the sum of two terms, one for each possible emitted photon helicity. Everything in the second line can be mapped directly to what would be obtained in a plane wave background  -- it may not look familiar when comparing to the literature because in the plane wave case one usually eliminates $q_\mu$ in favour of $\ell_\mu$, whereas here we have lost conservation of transverse momentum. Recall that this is encoded in the Gaussian factor $\triangle_\LCperp$, which also contains the usual `memory'~\cite{Dinu:2012tj,Bieri:2013hqa} term coming from $f_\infty$. (The argument of $\triangle_\LCperp$ may equivalently be written $(\ell+\sfpi_q-\sfpi_p)_\LCperp$.) The lack of transverse momentum conservation also impacts longitudinal momentum conservation through (\ref{def:qstar}) -- in the plane wave case the delta function would instead set $(\ell+q-p)_\LCp=0$.  

\subsection{All-multiplicity results}\label{sec:amplitude-transform}
It is clear that the contour deformations used in the amplitude calculation above are effectively \emph{undoing} the conformal transformation from $x^\mu$ to $\sigma^\mu$. This brings the amplitude into a form which is very closely related to that in a plane-wave background but not, crucially, equal to it -- there are focussing effects. In this section we therefore use the conformal transformation to reduce amplitudes for an \emph{arbitrary} number of photon emissions to what are essentially plane wave calculations. We first present the amplitude, then consider how its constituent parts behave under conformal transformation, before combining those parts and obtaining the result.

\medskip

\paragraph*{Amplitude:}
We again consider the tree-level scattering of an electron, momentum $p$ to momentum $q$, on the flying focus background with total depletion, but now with the emission of $N$ photons, momenta $\ell_j$ and polarisations $\varepsilon_j\equiv  \varepsilon^{(h_j)}(\ell_j)$. The amplitude is constructed from the sum over all possible permutations of the three-point and four-point vertices. A general term in this sum has the form
\be\label{N-point-amp1}
\begin{split}
    \mathcal{A}_{N} = (-ie)^N \, \e^{W[\alpha]}\,\int\!\ud^4x_1&\cdots \ud^4x_N\,  {\bar\Phi}_{q}(x_n) \,\e^{i\ell_n\cdot x_n}\cdots \e^{i\ell_1\cdot x_1} \\
&\cdots G(x_{j+1},x_{j})\varepsilon_j\cdot \overset{\leftrightarrow}{D}(x_j) G(x_j,x_{j-1})
\cdots 
G(x_{r+1},x_{r})\varepsilon_r\cdot\varepsilon_{r-1} G(x_r,x_{r-1}) \cdots \Phi_p(x_1) \;,
\end{split}
\ee
in which we have chosen some set of labels for the photons, $G(x,y)$ is the scalar propagator in the flying focus background, and each three and four-point vertex may be repeated multiple times. (These vertices may also appear next to the incoming and outgoing electron wavefunctions.) We take the amplitude and perform the complex conformal transformation $x\to \sigma$ as defined in (\ref{compConfTrans}). Recalling the definition of the conformal weight $\Omega=1+i\,k\,x^\LCp = 1/(1-i\,k\,\sigma^\LCp)$, the integration measures transform as
\be\label{measure-rule}
    \ud^4 x_i = \ud^4 \sigma_i\ \Omega^4(\sigma_i) \;,
\ee
and we can now consider the transformation of the various other components in (\ref{N-point-amp1}). 

\medskip

\paragraph*{Wavefunctions and propagator:}

As established in Sec.~\ref{sec:phi-from-conformal}, the scalar wavefunction obeys
\be
    \Phi(x) = \Omega^{-1} \phi(\sigma) 
\ee
in which $\phi$ is the Volkov solution in a plane wave background. The new ingredient we must consider in (\ref{N-point-amp1}) is the (scalar) electron propagator in the flying focus background. A massless Volkov solution can easily be extended off-shell by replacing the massless 4-momentum $p_\mu$ with a massive 4-momentum (i.e., $p^2\not=0$); a propagator is then easily constructed by stitching together these off-shell solutions. However, this strategy does not immediately extend to the flying focus background: the mass scale introduced in this way breaks conformal invariance and the resulting wavefunctions are not off-shell in a controlled way. Nevertheless, the conformal transformation linking plane wave and flying focus backgrounds can still be exploited to arrive at a propagator as follows.

We know that the propagator obeys, for an arbitrary function $J(x)$, 
\be\label{Gdef}
    D^2(x) \int\!\ud^4x'\, G(x,x') J(x') =  J(x) \;,
\ee
and that under conformal transformation the covariant derivative obeys (\ref{D2-transform}), which we repeat here:
\be\label{cov-transform}
    D^2(x) = \Omega^{-3} D^2_{\text{pw}}(\sigma) \Omega \;.
\ee
Using (\ref{cov-transform}) in (\ref{Gdef}), and also performing the conformal change of variable in the integral, $x'\to \sigma'$, we quickly land on
\be
    D^2_{\text{pw}}(\sigma)  \int\!\ud^4 \sigma'
    \bigg[\frac{G(\sigma,\sigma')}{(1-i k \sigma^{\LCp})(1-i k \sigma^{\prime \LCp})}\Bigg] j(\sigma') = j(\sigma) \;,
\ee
in which we have written $j(\sigma)\equiv J(\sigma)/(1-i\,k\,\sigma^\LCp)^3$, which is still arbitrary. It follows that the object in square brackets above \emph{is} the Green function in a (totally depleting) plane wave, call it $G_{\text{pw}}$, which is, of course, known exactly. Hence, as long as the propagator $G$ appears under at least one integral, which it always does in the amplitude (\ref{N-point-amp1}), then we may write it in terms of the plane wave propagator as
\be\label{propagator-result}
    G(x_i,x_j)
    = (1-i k \sigma_i^\LCp) G_{\text{pw}}(\sigma_i,\sigma_j)(1-i k \sigma_j^\LCp)
    = \Omega^{-1}(\sigma_i) G_{\text{pw}}(\sigma_i,\sigma_j) \Omega^{-1}(\sigma_j)\;.
\ee
As a consistency check, observe that the propagator transforms as a product of wavefunctions, as expected.

\medskip

\paragraph*{Vertices:} We begin with the three-point vertex. Spatial derivatives in the vertex always appear coupled to an external photon polarisation; changing variables we have, for an arbitrary scalar function $F(x)$,
\be
    \varepsilon\cdot \frac{\partial}{\partial x}  \frac{1}{1+i\,k\, x^\LCp}F(x)
    =
    \varepsilon^\nu \frac{\partial \sigma^\mu}{\partial x^\nu} \frac{\partial}{\partial \sigma^\mu} (1-i\,k\, \sigma^\LCp) F(\sigma) \;.
\ee
Now, since we use (anti)-lightfront gauge $\varepsilon^\LCp=0$, see (\ref{photonpol}), there is no $\partial/\partial \sigma^\LCp$ term in the above, hence the final $\sigma^\LCp$--dependent factor passes through the derivative. From here it is easily checked  that, for the full covariant derivative, 
 \be\label{three-point-rule}
 \varepsilon\cdot D(x) \,
 \Omega^{-1}(x)F(x) = \Omega^{-3}(\sigma) \mathcal{E}(\ell;\sigma)\cdot D_{\text{pw}}(\sigma)F(\sigma)\,,
\ee
in which we have \emph{defined}
\be\label{dressed-pol-def}
\begin{split}
    \mathcal{E}^\mu(\ell;\sigma) &:= \Omega^2(\sigma)
    \varepsilon^\nu(\ell) \frac{\partial \sigma^\mu}{\partial x^\nu} \\
   & =\frac{1}{\sqrt{2}\,(1-i\,k\,\sigma^\LCp)} \,\bigg(\frac{\ell_1 + i\, h\, \ell_2 + i\,k \, \ell_\LCm (\sigma^1+i\, h\, \sigma^2)}{\ell_\LCm\,(1- i\,k\,\sigma^\LCp)},\, 0,\, 1,\, i\, h\bigg)\;.
\end{split}
\ee
The reason for this definition will become clear in a moment (but compare (\ref{photonpol})). As for the contribution from four-point vertices, observe that, in terms of (\ref{dressed-pol-def}), we may write
\be\label{four-point-rule}
    \varepsilon_{i}\cdot \varepsilon_{j}
    =
    \Omega^{-1}(\sigma)\, 
    \mathcal{E}(\ell_i;\sigma) \cdot \mathcal{E}(\ell_j;\sigma)\,
    \Omega^{-1}(\sigma) \;,
\ee
in which $\sigma$ is the appropriate vertex position (labelled $\sigma_r$ in the example contribution (\ref{N-point-amp1})). We will use this result immediately below.

\medskip

\paragraph*{Assembly:}
Putting everything together, our change of variables turns wavefunctions, propagators and covariant derivatives into their plane wave counterparts. Each wavefunction is accompanied by a factor of $\Omega^{-1}$, each propagator by two such factors (one for each argument).
    At each four-point vertex (\ref{four-point-rule}), we generate two further such factors, for a total of four. These are cancelled exactly by the factor of $\Omega^4$ in the measure (\ref{measure-rule}). At each three-point vertex we obtain three powers of $\Omega^{-1}$ from the covariant derivative acting on a propagator/wavefunction, see (\ref{three-point-rule}), and one power of $\Omega^{-1}$ from the propagator/wavefunction which multiplies this. This is again cancelled by the measure at this vertex. Hence all factors of $\Omega$ cancel from the amplitude and we arrive at 
\begin{align}\label{N-point-amp1-transformed}
    \mathcal{A}_{N} &= (-ie)^N \, \e^{W[\alpha]}\,\int\!\ud^4\sigma_1\cdots \ud^4\sigma_N\,  {\bar\phi}_{q}(\sigma_n) \,
    \e^{i\ell_n\cdot x(\sigma_n)}\cdots \e^{i\ell_1\cdot x(\sigma_1)} \\
&\cdots G_{\text{pw}}(\sigma_{j+1},\sigma_{j})\mathcal{E}(\ell_j;\sigma_j)\cdot {\overset{\leftrightarrow}{D}}_{\text{pw}}(\sigma_j) G_{\text{pw}}(\sigma_j,\sigma_{j-1})
\cdots 
G_{\text{pw}}(\sigma_{r+1},\sigma_{r})\mathcal{E}(\ell_{r};\sigma_r)\cdot\mathcal{E}(\ell_{r-1};\sigma_r) G_{\text{pw}}(\sigma_r,\sigma_{r-1}) \cdots \phi_p(\sigma_1) \;, \nonumber
\end{align}
in which we note the transformed exponents in the emitted photon wavefunctions, and the new functions $\mathcal{E}$ replacing the usual photon polarisations.

We have thus reduced the $N$-photon emission amplitude in a flying focus background to, essentially, that in a plane wave background -- all wavefunctions and scalar propagators are given by the usual Volkov expressions, the flying focus field is replaced by its plane wave counterpart. The only difference compared to the plane wave amplitude is that the usual free wavefunctions for emitted photons $\varepsilon^\mu(\ell)\, \e^{i\ell\cdot x}$ are replaced by new functions $\mathcal{E}^\mu(\ell;\sigma)\, \e^{i\ell\cdot x(\sigma)}$ -- all the complexity of beam focussing has been shifted into these functions. This is interesting in itself, but there is an even simpler representation of this result which also lends itself to explicit evaluation of amplitudes, as we shall now see.
%%%%%%%%%%%%%%%%%%%%%%%%%%%%%%%
\subsection{Focussing for free}
%%%%%%%%%%%%%%%%%%%%%%%%%%%%%%%
It is straightforward to verify that the new functions appearing in our amplitude obey 
\be
    \partial \cdot \big(\mathcal{E}(\sigma)e^{i\ell\cdot x(\sigma)}\big) = 0 \;,
    \qquad
    \text{and}
    \qquad 
    \partial^2 \big(\mathcal{E}^\mu(\sigma)\,
    \e^{i\ell\cdot x(\sigma)}\big) = 0  \;,
\ee
and so are solutions of the vacuum Maxwell equations. In other words, the new functions \emph{are} wavepackets of free photons -- performing the conformal transformation on the flying focus amplitude has in fact converted it into the equivalent plane wave amplitude, but with the usual (plane wave) asymptotic photon states replaced by photon wavepacket states. We have genuinely reduced the flying focus calculation to one in a plane wave background.

It will at this stage be unsurprising to learn that the wavepacket in question is nothing but the same Gaussian appearing in the definition of the flying focus beam itself:
\be\label{ConfTransPhoton}
\begin{split}
    \mathcal{E}^\mu (\ell;\sigma)\,\e^{i \ell\cdot x(\sigma)}
    &=
    \int_q 2 q_\LCm\,
    \hat\delta(q_\LCm-\ell_\LCm)
    \bigg[\frac{2\pi}{k \ell_\LCm}\,
    \e^{-|q_\LCperp-\ell_\LCperp|^2/(2 k \ell_\LCm)}\bigg]
    \, \varepsilon^\mu(q)\,
    \e^{i q_\mu \sigma^\mu} \\
    &=: \langle\!\langle\, \varepsilon^\mu(\ell) \, \e^{i\ell\cdot \sigma}\,\rangle\!\rangle \,,
\end{split}
\ee
which defines the Gaussian average $\langle\!\langle \cdot \rangle\!\rangle$. We therefore have the remarkable final result that beam focussing effects can be added to total depletion plane wave tree-level amplitudes simply by taking a Gaussian average over the emitted photon momenta:
\be\label{ConfTransAmpFinal}
    \mathcal{A}_N(p\to q + \ell_1 +\cdots+ \ell_N)
    =
    \e^{W[\alpha]}\,\big\langle\!\big\langle\, \mathcal{A}^\text{pw}_N(p\to q + \ell_1 +\cdots+ \ell_N)\,\big\rangle\!\big\rangle \;,
\ee
in which there is an independent averaging for each photon. We stress that this result holds at arbitrary multiplicity.

%%%%%%%%%%%%%%%%%
%%%%%%%%%%%%%%%%%

\section{Partial depletion}\label{sec:partial-depletion}

Up to this point, we have concentrated on scattering in complex backgrounds which correspond to total depletion of an incoming coherent state. However, as reviewed in Sec.~\ref{sec:review}, partial depletion can also be modelled by scattering on complex backgrounds which have both incoming and outgoing parts. While this is trivial to accommodate for plane wave backgrounds, it is far more challenging in flying focus backgrounds -- at least, as they have been constructed here. This is because one needs to use \emph{different} conformal transformations for the incoming and outgoing parts of the field, and it is not clear how to construct massless wavefunctions in the resulting background.

To tackle this problem, we restrict our attention in this section to the setting where both the initial and final coherent states collectively give \emph{anti-self-dual} (ASD) fields. This corresponds to taking $\alpha_-=\beta_+=0$, so that all incoming coherent state photons (encoded in $\alpha_+$) are positive helicity and all outgoing coherent state photons (encoded in $\beta_-$) are negative helicity. While ASD fields are certainly not generic physical configurations for partial depletion, they do allow for significant computational simplifications. For instance, the ASD assumption has enabled the determination of exact wavefunctions, high-loop and high-multiplicity scattering amplitudes for backgrounds in both gauge theory and gravity~\cite{Dunne:2001pp,Dunne:2002qf,Dunne:2002qg,Adamo:2020syc,Adamo:2020yzi,Adamo:2022mev,Costello:2022jpg,Bogna:2023bbd,Adamo:2023fbj,Guevara:2023wlr,Adamo:2024xpc,Garner:2024tis,Guevara:2024edh,Bittleston:2024efo,Adamo:2025fqt,Bittleston:2026pai}, and provided insights into the double copy~\cite{Brown:2023zxm,Kim:2024dxo,Ilderton:2025gug} between gauge theories and gravity, $\beta$-functions~\cite{Bittleston:2025jmk} and loop-level scattering amplitudes in QCD~\cite{Costello:2023vyy,Dixon:2024mzh,Dixon:2024tsb,Morales:2025alm,Badger:2025uym}.

Below, we provide a brief overview of ASD backgrounds which are totally or only partially depleting. We then observe that in the case of total depletion, there is an alternative basis of massless wavefunctions, compared to those studied above. We show that these can be generalised to ASD, partially depleting flying focus backgrounds, providing a toolkit to go beyond total depletion.

%%%%%%%%%%%%%%%%%%%%%%

\subsection{Anti-self-dual fields}

An electromagnetic field $F_{\mu\nu}$ is said to be \emph{ASD} if it obeys
\be\label{SDdefn}
{\tilde F}_{\mu\nu} \equiv \frac{1}{2}\,\epsilon_{\mu\nu\rho\sigma}\,F^{\rho\sigma}=-i\,F_{\mu\nu}\,.
\ee
Anti-self-duality is a powerful feature: ASD field strengths automatically solve the vacuum Maxwell equations (by virtue of the Bianchi identity) and the ASD condition gives a set of first-order partial differential equations for the gauge potential which are classically integrable~\cite{Ward:1977ta,Belavin:1978pa}.

Thanks to the factor of $i$ appearing in~\eqref{SDdefn}, all ASD fields are inherently complex-valued. Following the logic of Sec.~\ref{sec:complexbackgrounds}, these will still represent real scattering processes when the ASD fields encode incoming and outgoing coherent states of photons. These correspond to gauge potentials of the form
\begin{equation}\label{SDcohstate}
    A_{\mu}(x)=
\int_{\ell}\, \varepsilon_{\mu}^{(+)}\left[\alpha_{+}(\ell)\,\e^{-i\,\ell\cdot x}+\beta^{*}_{-}(\ell)\,\e^{i\,\ell\cdot x}\right]\;.
\end{equation}
In other words, scattering in the ASD background~\eqref{SDcohstate} encodes transitions from an incoming coherent state of positive helicity photons $\alpha_+$ to an outgoing coherent state of negative helicity photons $\beta_-$.

When dealing with ASD plane waves and flying focus solutions, it is convenient to work in coordinates where the transverse $x^\perp$--plane is identified with the complex plane. If $(z,\bar{z})$ are complex conjugate coordinates on the complex plane, then the Minkowski metric becomes
\be\label{CMinkmet}
z=\frac{x^1+i\,x^2}{\sqrt{2}}\,, \quad \bar{z}=\frac{x^1-i\,x^2}{\sqrt{2}}\,, \qquad \ud s^2=2\left(\ud x^\LCm\,\ud x^+-|\ud z|^2\right)\,.
\ee
In these coordinates, the profile functions $f_{\perp}(x^-)$ can be combined into complex functions
\be\label{Compprofile}
\sff \equiv \frac{f_1-i\, f_2}{\sqrt{2}}\,, \qquad \bar\sff \equiv \frac{f_1 +i\, f_2}{\sqrt{2}}\,,
\ee
so that the plane wave gauge potential \eqref{PlaneWavepot} becomes
\be\label{PWcomplex1}
A_{\mathrm{pw}}(x)=-\left[z\,\dot{\sff}(x^-)+\bar{z}\,\dot{\bar\sff}(x^-)\right] \ud x^{\LCm}\,.
\ee
For complex-valued fields, $\sff$ and $\bar\sff$ are \emph{not} related by complex conjugation, but are instead independent functions -- an ASD plane wave corresponds to $\bar\sff=0$ (or $f_1=-i\,f_2$). In this case, the gauge field has a single component in either of the gauges used in Sec.~\ref{sec:build-FF}, and we will work with:
\be\label{SDPW1}
A_{\mathrm{asd}}=\sff(x^-)\,\ud z
\quad
\implies
\quad
F_{\mathrm{asd}}=\dot{\sff}(x^-)\,\ud x^{\LCm}\wedge\ud z\,,
\ee
which is easily seen to obey~\eqref{SDdefn}. In the correspondence with coherent state scattering, the positive (negative) frequency modes of the profile function $\sff$ correspond to the incoming (outgoing) coherent state $\alpha_+$ ($\beta_-$).

Just as we obtained totally-depleting flying focus solutions from a (complex) conformal transformation of a totally depleting plane wave, one imagines that something similar occurs in the ASD but partially depleting case. However, this expectation is too na\"ive: after performing the conformal transformation \eqref{compConfTrans}, the resulting gauge potential will be proportional to $\sff(\sigma^-)$. Since $\sff$ has both positive and negative frequency Fourier modes, the latter will produce exponential growth in the transverse $(z,\bar{z})$-plane, rather than the desired exponential decay, as discussed in Sec.~\ref{sec:build-FF}.

One resolution to this problem is to use two distinct conformal transformations: one on the incoming modes of $\sff$ and another on the outgoing modes.  Specifically, let
\be\label{Fourierdecomp}
\begin{split}
\sff(x^-)&=\int_{0}^{\infty}\frac{\ud \omega}{2\pi}\,\sff_{+}(\omega)\,\e^{-i\,\omega\,x^-}+\int_{0}^{\infty}\frac{\ud \omega}{2\pi}\,\sff_{-}(-\omega)\,\e^{i\,\omega\,x^-} \\
 &=\sff_{+}(x^-)+\sff_{-}(x^-)\,,
\end{split}
\ee
be the decomposition of $\sff$ into its incoming/positive frequency ($\sff_{+}$) and outgoing/negative frequency ($\sff_-$) parts. Then we act with the conformal transformation \eqref{compConfTrans} on the positive frequency part of the plane wave, while acting on the negative frequency part with
\be\label{compConfTrans2}
x^{\mu}\to\tilde{\sigma}^{\mu}\,, \qquad \tilde{\sigma}^-=x^-+i\,\tilde{k}\,\frac{|z|^2}{1-i\,\tilde{k}\,x^+}\,, \quad \tilde{\sigma}=\frac{z}{1-i\,\tilde{k}\,x^+}\,, \quad \tilde{\sigma}^+=\frac{x^+}{1-i\,\tilde{k}\,x^+}\,,
\ee
where $\tilde{k}\geq0$. The resulting gauge potential is
\be\label{ND-SDFF}
A=\frac{\sff_+(\sigma^-)}{1+i\,k\,x^+}\left(\ud z-\frac{i\,k\,z}{1+i\,k\,x^+}\,\ud x^+\right)+\frac{\sff_-(\tilde{\sigma}^-)}{1-i\,\tilde{k}\,x^+}\left(\ud z+\frac{i\,\tilde{k}\,z}{1-i\,\tilde{k}\,x^+}\,\ud x^{\LCp}\right)\;,
\ee
whose incoming Fourier modes decay as before, while the outgoing Fourier modes now also decay exponentially due to the sign change in (\ref{compConfTrans2}). Thus~\eqref{ND-SDFF} is a partially depleting flying focus field, where the focussing of the incoming field is controlled by $k$, while the focussing of the outgoing field is controlled by $\tilde{k}$. It is easy to check that the corresponding field strength is ASD, as expected since the ASD condition~\eqref{SDdefn} is conformally invariant. 

%%%%%%%%%%%%%%%%%%%%%%%%%%%%%%%%%%%%%%%%

\subsection{Alternative wavefunctions for anti-self-dual total depletion}

To build a toolkit for computing amplitudes on the partial depleting ASD flying focus background \eqref{ND-SDFF}, we need explicit wavefunctions for charged particles on the background. Recall that in the totally depleting case, we obtained wavefunctions in the flying focus background by applying the conformal transformation \eqref{compConfTrans} to the Volkov wavefunctions of the plane wave background. In the partially depleting case, we have \emph{two} independent conformal transformations -- one for the incoming and one for the outgoing field modes -- and it is not clear which one to apply to the Volkov solutions.

To resolve this problem, we first return to the setting of total depletion: the incoming ASD flying focus background is given by 
\be\label{TD-SDFF}
A=\frac{\sff_+(\sigma^-)}{1+i\,k\,x^+}\left(\ud z-\frac{i\,k\,z}{1+i\,k\,x^+}\,\ud x^+\right)\,,
\ee
with $\sff_+$ having only positive frequency Fourier modes. In this case, we can simply adapt our previous solutions for massless charged scalar wavefunctions to the special case of an ASD background:
\be\label{TD-SDwfs}
\Phi_p(x)=\frac{1}{1+i\,k\,x^+}\,\exp\!\left[-i\,p\cdot\sigma
+ \frac{i\,e\,\mathsf{p}}{p_+}\,\int^{\sigma^-}\!\!\ud s\,\sff_+(s)\right]\,,
\ee
where, here and below, $\mathsf{p}:=(p_1+i\,p_2)/\sqrt{2}$.

As we have seen, these wavefunctions are well-adapted to performing explicit computations of amplitudes in totally-depleting flying focus fields. Furthermore, when $k\to0$ they reduce to the Volkov solutions \eqref{Volkov} in an ASD (totally depleting) plane wave. However, their behaviour in the free limit is somewhat non-standard: considering the $e\to0$ limit where the scalar decouples from the background, the wavefunction \eqref{TD-SDwfs} goes to
\be\label{e20.1}
\begin{split}
\lim_{e\to0}\Phi_p(x)&=\frac{1}{1+i\,k\,x^+}\,\e^{-i\,p\cdot\sigma} \\
& =\frac{2\pi}{k\,p_-}\,\int\hat{\ud}^{2}q_{\perp}\,\exp\!\left[-\frac{|q_\perp-p_{\perp}|^2}{2k\,p_-}-i\,q_\perp\,x^\perp-i\,\frac{|q_\perp|^2}{2\,p_-}\,x^+-i\,p_{-}\,x^{-}\right]\,.
\end{split}
\ee
In particular, we see that $\Phi_p$ goes to a Gaussian wavepacket of free plane waves as $e\to 0$. 

While there is nothing wrong with this (such Gaussian wavepackets are simply an over-complete basis of solutions in the free limit), one might have hoped that there exist wavefunctions which reduce to a (single) free plane wave in the $e\to 0$ limit. Remarkably, such solutions can be obtained by simply solving the Klein-Gordon equation order-by-order in $e$ starting from $\e^{-ip\cdot x}$. The resulting power series can be re-summed to give the alternative wavefunctions
\be\label{pertWF1}
\psi_p(x)=\exp\!\left[-i\,p\cdot x
+
\frac{i\,e\,\mathsf{p}}{p_+\,(1+i\,k\,x^+)+i\,k\,\mathsf{p}\,\bar{z}}\,\int_{\tau}^{\sigma^-}\!\!\ud s\,\sff_{+}(s)\right]\,,
\ee
where the lower limit of integration appearing in the exponential is
\be\label{tau-def}
\tau\equiv \tau(x) = \frac{p\cdot x}{p_-}-i\,\frac{p_+}{k\,p_-}\,.
\ee
Clearly, as $e\to0$, $\psi_p(x)\to\e^{-i p\cdot x}$, as desired. The appearance of the quantity $\tau(x)$ as a lower limit of integration may seem surprising, but it plays two important roles. First, it ensures that $\psi_p(x)$ is non-singular  -- when the denominator in (\ref{pertWF1}) vanishes, it can be checked that $\tau \to \sigma^\LCm$, pinching the integral to zero such that the exponent remains finite. Furthermore, the imaginary part of $\tau(x)$ provides a Gaussian damping factor in the $k\to 0$ limit, ensuring that we recover the Volkov solution.

%%%%%%%%%%%%%%%%%%%%%%%%%%%%%%%%%%

\subsection{Wavefunctions for partial depletion}

The alternative, perturbative basis of wavefunctions $\psi_p$ can now be generalised to provide exact wavefunctions in the ASD flying focus background \eqref{ND-SDFF} with incoming \emph{and} outgoing parts. One simply takes
\be\label{PD-wfns}
\psi_p(x)=\exp\!\left[-i\,p\cdot x
+
\frac{i\,e\,\mathsf{p}}{p_+\,(1+i\,k\,x^+)+i\,k\,\mathsf{p}\,\bar{z}}\,\int_{\tau}^{\sigma^-}\!\!\ud s\,\sff_{+}(s)
+
\frac{i\,e\,\mathsf{p}}{p_+\,(1-i\,\tilde{k}\,x^+)-i\,\tilde{k}\,\mathsf{p}\,\bar{z}}\,\int^{\tilde{\sigma}^-}_{\tilde{\tau}}\!\!\ud s\,\sff_-(s)\right]\,,
\ee
in which
\be\label{tau-tilde-def}
    \tilde\sigma = \sigma|_{k\to -\tilde k} 
    \qquad\text{and}\qquad 
    \tilde\tau = \tau|_{k\to -\tilde k} \;.
\ee
This $\psi_p(x)$ is essentially the product of solutions in each of the positive and negative frequency fields. It can be verified by direct calculation that (\ref{PD-wfns}) solves the massless Klein-Gordon equation in the ASD flying focus background (\ref{ND-SDFF}) -- see the Appendix for details. Furthermore, these solutions reduce to free plane waves as $e\to 0$, are non-singular, and produce Volkov solutions coupling to the positive/negative frequency modes of the background when focussing is turned off ($k/\tilde{k}\to0$).

In principle, these wavefunctions provide us with the tools needed to compute processes like non-linear Compton scattering in a partially-depleted, focussed background. Unfortunately, so far we have been unable to find a way to analytically evaluate the integrals which arise in amplitude calculations using the $\psi_p$ wavefunctions. In essence, our ability to perform calculations in the totally depleting case followed from using the wavefunctions $\Phi_p$ generated by a conformal transformation: the contour prescriptions or Gaussian averaging used to evaluate the resulting integrals essentially followed from this conformal transformation acting on the usual integrals in plane wave backgrounds. Of course, it may still prove to be possible to perform analytic calculations with the wavefunctions $\psi_p$. In any case, the integrals arising in amplitudes evaluated on these wavefunctions should be amenable to numerical integration.

We note, finally, an interesting consequence of the fact that our partially-depleting field model contains both incoming and outgoing parts -- the field is no longer null:
\be\label{non-null}
    F_{\mu\nu}\,F^{\mu\nu} = i\, F_{\mu\nu}{\tilde F}^{\mu\nu}
    =  -4
    \frac{z^2\, {\dot \sff}_\LCp(\sigma^\LCm)\,  {\dot \sff}_\LCm({\tilde \sigma}^\LCm)}{(1+i\,k\, {x^\LCp})^3 (1-i\, {\tilde k}\, {x^\LCp})^3}\, (k+{\tilde k})^2 \;.
\ee
In a real field, non-zero invariants can signal e.g.~the possibility of spontaneous particle production from the vacuum. The interpretation here, though, is not so clear. Working perturbatively, one can easily check that the leading contribution to the $0\to 2$ process, which goes like $e^2\alpha^2$ for any initial coherent state with profile $\alpha$, vanishes because both incoming photons have the same helicity. Similarly, the order $e^3\alpha^2\beta$ contribution vanishes. These are of course examples of `all plus' (helicity) amplitudes which vanish at all multiplicity~\cite{Gastmans:1990xh}, suggesting that the perturbative contribution to the $0\to2$ amplitude in an ASD flying focus is zero. It would be very interesting to investigate particle creation processes in more detail, potentially using the wavefunctions (\ref{PD-wfns}).

%%%%%%%%%%%%%%%%%%%%%%%%%%%%%%%
%%%%%%%%%%%%%%%%%%%%%%%%%%%%%%%

\section{Conclusions}\label{sec:conclusions}

We have investigated strong-field QED scattering processes in a class of highly inhomogeneous backgrounds which includes flying focus beams. We have shown that these fields can be obtained from conformal transformations of plane waves, and that the corresponding wavefunctions describing massless scalars scattering on the backgrounds are given by applying the same transformation to the Volkov wavefunctions.

While the required conformal transformation, and resulting electromagnetic fields, were complex, the well-known equivalence between backgrounds and coherent states meant that we were still able to calculate sensible amplitudes, namely those in which the flying focus beam was completely absorbed during some scattering process. We considered the amplitudes for $N$-photon emission from an electron scattering on the background and, by exploiting the conformal transformation, showed that this was given by a simple Gaussian average over the emitted photon momenta in the corresponding plane wave amplitude. This is an elegant result which shows how much-sought-after focussing effects can be included into plane-wave calculations with little additional effort, and furthermore adds to the library of results which show how to use plane wave amplitudes to construct amplitudes in other backgrounds~\cite{Adamo:2021jxz,Copinger:2024twl,Ilderton:2024ufp,Copinger:2026sjy}.

Finally, we considered the extension of our results to partial depletion amplitudes.
In this case it is unclear how to construct the corresponding massless charge wavefunctions via a conformal transformation. To make progress, we made the simplifying assumption that the flying focus background was anti-self-dual; physically, this means that all incoming coherent state photons are either absorbed or, if not, flip helicity.
In this setting, we were able to find a set of charge wavefunctions with an intriguing structure, seemingly quite different to that encountered earlier in the paper. The continued investigation of these new solutions is a natural target for future work, including their use in calculating amplitudes and, if possible, extending the solutions beyond the anti-self-dual sector.

It would also be interesting to see whether the connection between amplitudes in plane wave and flying focus backgrounds, summarised by the averaging result (\ref{ConfTransAmpFinal}), can be generalised to scattering in Yang-Mills and gravitational backgrounds or, even in QED, to loop corrections. Regarding the latter, we do not expect renormalisation effects to impact the averaging result, at least to one-loop, but it is possible that constraints may appear which limit the result to e.g.~special kinematics -- this is what happens to similar results linking scattering in shockwave and plane wave backgrounds when going from tree-level to one-loop~\cite{Ilderton:2024ufp}. Finally, it would also be interesting to seek other, non-conformal, transformations which allow us to `lift' plane wave results to other background fields of interest, as we have done here for the flying focus. 

\medskip

\textit{We thank Alex Goodenbour, Richard Myers and Karthik Rajeev for interesting conversations. The authors are supported by the STFC grant UKRI1887
(AN), the STFC consolidated grant ``Particle Theory at the Higgs Centre" ST/X000494/1 (TA, AI), a Royal Society University Research Fellowship (TA), the Simons Collaboration on Celestial Holography MPS-CH-00001550-11 (TA), and the ERC Consolidator/UKRI Frontier grant ``TwistorQFT'' EP/Z000157/1 (TA).}

%%%%%%%%%%%%%%%%%%%%%%%%%%%%
%%%%%%%%%%%%%%%%%%%%%%%%%%%%

\appendix

\section{Properties of wavefunctions for partial depletion}

In this appendix, we provide some further detail on the wavefunctions \eqref{PD-wfns} describing massless charged scalars in the anti-self-dual, partially-depleting flying focus background \eqref{ND-SDFF}. A direct, brute force calculation verifies that these wavefunctions are indeed solutions, but here we give a more illustrative explanation of this fact. 

Let us denote the solutions by $\psi_{p}(x)=\e^{-i\mathcal{S}_{p}(x)}$, where, from~(\ref{PD-wfns}) and using the notation in (\ref{tau-def}) and (\ref{tau-tilde-def}),
\be\label{app:WKBexp}
\mathcal{S}_{p}(x):=p\cdot x-
\frac{e\,\mathsf{p}}{p_+\,(1+i\,k\,x^+)+i\,k\,\mathsf{p}\,\bar{z}}\,\int_{\tau}^{\sigma^-}\!\!\ud s\,\sff_{+}(s)-
\frac{e\,\mathsf{p}}{p_+\,(1-i\,\tilde{k}\,x^+)-i\,\tilde{k}\,\mathsf{p}\,\bar{z}}\,\int^{\tilde{\sigma}^-}_{\tilde\tau}\!\!\ud s\,\sff_-(s)\,.
\ee
We also define $\Pi_{\mu}:=\partial_{\mu}\mathcal{S}_{p}-e\,A_{\mu}$, where $A_{\mu}$ is the background Maxwell field \eqref{ND-SDFF}. With this notation, the massless Klein-Gordon equation becomes
\be\label{app:wave-eq}
D^2\psi_{p}=-\psi_{p}\left(\Pi^2+i\,\partial\cdot\Pi\right)=0\,.
\ee
In fact, each term here must vanish independently for the equation to be solved (as can be seen by taking $\mathcal{S}_{p}\to\mathcal{S}_p/\hbar$), so solving the massless wave equation in the background is equivalent to solving
\be\label{app:wave-eq2}
\Pi^2=0=\partial\cdot\Pi\,,
\ee
so that the wavefunction is WKB exact. In other words, we must show that the `dressed momentum' $\Pi_\mu$ associated to $\psi_{p}$ is null, and has vanishing divergence.

The characterisation of null vectors in four-dimensional Minkowski spacetime is particularly transparent in the 2-spinor, or spinor helicity, formalism, where vector indices are traded for a pair of SL$(2,\mathbb{C})$ Weyl spinor indices (cf. \cite{Penrose:1985bww,Srednicki:2007qs,Elvang:2013cua,Dixon:2013uaa,Cheung:2017pzi}). This formalism has been extensively used in the study of massless scattering in trivial backgrounds, but it also extends to scattering in the presence of background fields~\cite{Adamo:2019zmk,Adamo:2020qru}. The translation between standard vector indices and 2-spinor indices is accomplished by contraction with the Pauli matrices; if $x^{\mu}=(x^0,x^1,x^2,x^3)$ is the coordinate 4-vector then its 2-spinor representation is given by the $2\times 2$ matrix
\be\label{app:2spinorcon}
x^{\alpha\dot\alpha}:=x^{\mu}\,\sigma_{\mu}^{\alpha\dot\alpha}=\left(\begin{array}{cc}
x^+ & \bar{z} \\
z & x^-
\end{array}\right)\,,
\ee
where $\alpha,\,\dot\alpha=0,1$ are SL$(2,\mathbb{C})$ Weyl spinor indices of negative/positive chirality and $\sigma_{\mu}^{\alpha\dot\alpha}$ are the Pauli matrices. The Minkowsi metric \eqref{CMinkmet} becomes
\be\label{app:Minkmet}
\ud s^2=\epsilon_{\alpha\beta}\,\epsilon_{\dot\alpha\dot\beta}\,\ud x^{\alpha\dot\alpha}\,\ud x^{\beta\dot\beta}\,,
\ee
in these variables, where
\be\label{app:LeviCivita}
\epsilon_{\alpha\beta}=\left(\begin{array}{cc}
0 & 1 \\
-1 & 0
\end{array}\right)=\epsilon_{\dot\alpha\dot\beta}\,,
\ee
are the 2-dimensional Levi-Civita symbols.

Two-spinor indices are raised and lowered using these Levi-Civita symbols; we use the convention
\be\label{app:spinorconvention}
a^{\alpha}=\epsilon^{\alpha\beta}\,a_{\beta}\,, \qquad a_{\alpha}=a^{\beta}\,\epsilon_{\beta\alpha}\,,
\ee
and similarly for dotted spinors. It will also be useful to adopt some widely-used short-hand for Lorentz-invariant spinor contractions of each chirality
\be\label{app:spinorbrackets}
\la a\,b\ra:=a^{\alpha}\,b_{\alpha}\,, \qquad [c\,d]:=c^{\dot\alpha}\,d_{\dot\alpha}\,.
\ee
Finally, let $o_{\alpha}=(1,0)$ and $\iota_{\alpha}=(0,1)$ denote a spinor dyad, normalised so that $\la \iota\,o\ra=1$. 

A particularly powerful consequence of working in this 2-spinor formalism is that a 4-vector is null \emph{if and only if} its 2-spinor representation is simple -- that is, it is the product of two spinors of opposite chirality. For the null momentum $p^{\mu}$ associated to the wavefunction $\psi_p$, this means that 
\be\label{app:nullmom}
p^{\alpha\dot\alpha}=p^{\mu}\,\sigma_{\mu}^{\alpha\dot\alpha}=\kappa^{\alpha}\,\tilde{\kappa}^{\dot\alpha}
\ee
for some 2-spinors $\kappa^{\alpha}$, $\tilde{\kappa}^{\dot\alpha}$. For Lorentzian-real 4-momenta, these spinors are related by complex conjugation, $\tilde{\kappa}^{\dot\alpha}=\overline{\kappa^{\alpha}}$, but the following discussion holds for generic complex null momenta, so we will keep them independent. 

We can then proceed to calculate $\Pi_{\alpha\dot\alpha}$ directly, making use of the 2-spinor identities $p_{+}=\la\kappa\,\iota\ra\,[\tilde{\kappa}\,\iota]$, $\mathsf{p}=\la\kappa\,\iota\ra\,[o\,\tilde{\kappa}]$, 
\be\label{app:sigtau}
\partial_{\alpha\dot\alpha}\sigma^-=\iota_{\alpha}\,\iota_{\dot\alpha}-\frac{i\,k}{1+i\,k\,x^+}\left(o_{\alpha}\,\iota_{\dot\alpha}\,z+\iota_{\alpha}\,o_{\dot\alpha}\,\bar{z}\right)-\frac{k^2\,|z|^2}{(1+i\,k\,x^+)^2}\,o_{\alpha}\,o_{\dot\alpha}\,, \qquad \partial_{\alpha\dot\alpha}\,\tau=\frac{\kappa_{\alpha}\,\tilde{\kappa}_{\dot\alpha}}{\la\kappa\,o\ra\,[\tilde{\kappa}\,o]}\,,
\ee
where $\partial_{\alpha\dot\alpha}\equiv\frac{\partial}{\partial x^{\alpha\dot\alpha}}$, and the expression for the background field
\be\label{app:Maxwell}
A_{\alpha\dot\alpha}=\frac{\sff_{+}(\sigma^-)\,o_{\dot\alpha}}{1+i\,k\,x^+}\left(\iota_{\alpha}-\frac{i\,k\,z}{1+i\,k\,x^+}\,o_{\alpha}\right)+\frac{\sff_{-}(\tilde{\sigma}^-)\,
%-
o_{\dot\alpha}}{1-i\,\tilde{k}\,x^+}\left(\iota_{\alpha}+\frac{i\,\tilde{k}\,z}{1-i\,\tilde{k}\,x^+}\,o_{\alpha}\right)\,.
\ee
A straightforward, if somewhat tedious calculation, then reveals that $\Pi_{\alpha\dot\alpha}(x)=K_{\alpha}(x)\,\tilde{\kappa}_{\dot\alpha}$, for
\begin{multline}\label{app:K}
K_{\alpha}(x)=\kappa_{\alpha}\left(1+\frac{e\,\sff_{+}(\tau)}{(1+ikx^+)\,[\tilde{\kappa}\,\iota]+ik\,[o\,\tilde{\kappa}]\,\bar{z}}+\frac{e\,\sff_{-}({\tilde\tau})}{(1-i\tilde{k}x^+)\,[\tilde{\kappa}\,\iota]-i\tilde{k}\,[o\,\tilde{\kappa}]\,\bar{z}}\right) \\-\frac{i\,k\,e\,[o\,\tilde{\kappa}]\,o_{\alpha}}{[(1+ikx^+)\,[\tilde{\kappa}\,\iota]+ik\,[o\,\tilde{\kappa}]\,\bar{z}]^2}\,\int^{\sigma^-}_{\tau}\ud s\,\sff_{+}(s) 
+\frac{i\,\tilde{k}\,e\,[o\,\tilde{\kappa}]\,o_{\alpha}}{[(1-i\tilde{k}x^+)\,[\tilde{\kappa}\,\iota]-i\tilde{k}\,[o\,\tilde{\kappa}]\,\bar{z}]^2}\,\int^{\tilde{\sigma}^-}_{\tilde\tau}\ud s\,\sff_{-}(s) \\
+\frac{e\,\sff_{+}(\sigma^-)}{(1+ikx^+)\,[\tilde{\kappa}\,\iota]+ik\,[o\,\tilde{\kappa}]\,\bar{z}}\left(\iota_{\alpha}-\frac{i\,k\,z}{1+ikx^+}\,o_{\alpha}\right)+\frac{e\,\sff_{-}(\tilde{\sigma}^-)}{(1-i\tilde{k}x^+)\,[\tilde{\kappa}\,\iota]-i\tilde{k}\,[o\,\tilde{\kappa}]\,\bar{z}}\left(\iota_{\alpha}+\frac{i\,\tilde{k}\,z}{1-i\tilde{k}x^+}\,o_{\alpha}\right)\,,
\end{multline}
where the Schouten identity $\iota_{\dot\alpha}\,[o\,\tilde{\kappa}]+o_{\dot\alpha}\,[\tilde{\kappa}\,\iota]=-\tilde{\kappa}_{\dot\alpha}$ is repeatedly used. Of course, the expression for the spinor $K_{\alpha}(x)$ is quite complicated, but the crucial fact is that $\Pi_{\alpha\dot\alpha}$ is immediately seen to be null, as it is written as a product of 2-spinors. This establishes that $\Pi^2=0$, as required. Note that only the negative chirality (i.e., undotted) part of $\Pi_{\alpha\dot\alpha}$ is dressed by the background; the positive chirality part remains the un-dressed momentum spinor $\tilde{\kappa}_{\dot\alpha}$. This is an expected feature, reflecting the fact that the background is anti-self-dual~\cite{Adamo:2020syc,Adamo:2020yzi}.

To complete the proof that $\psi_{p}$ solves the wave equation, we must still establish that $\partial\cdot\Pi=0$, which is equivalent to 
\be\label{app:spinder}
\tilde{\kappa}^{\dot\alpha}\,\partial_{\alpha\dot\alpha}K^{\alpha}(x)=0\,.
\ee
Using the identities \eqref{app:sigtau} and the fact that spinor contractions are skew-symmetric (so that $\la a\,a\ra=0=[b\,b]$ for any 2-spinors $a_{\alpha}$ and $b_{\dot\alpha}$), it follows that only a few terms in the expression for $K^{\alpha}$ are non-vanishing on their own when acted upon with the differential operator $\tilde{\kappa}^{\dot\alpha}\partial_{\alpha\dot\alpha}$. These come from the action of the derivative on the third line of \eqref{app:K} as well as the upper limit of integration for terms in the second line. These contributions are then seen to cancel with each other: for the positive frequency terms one finds
\be\label{app:spinder2}
\frac{i\,k\,e\,\sff_+(\sigma^-)\,[\tilde{\kappa}\,o]}{(1+ikx^+)\,[(1+ikx^+)\,[\tilde{\kappa}\,\iota]+ik\,[o\,\tilde{\kappa}]\,\bar{z}]^2}\left[(1+ikx^+)\,[\tilde{\kappa}\,\iota]+ik\,\bar{z}\,[o\,\tilde{\kappa}]-(1+ikx^+)\,[\tilde{\kappa}\,\iota]+ik\,\bar{z}\,[\tilde{\kappa}\,o]\right]=0\,,
\ee
with a similar cancellation occurring for the contributions for negative frequency terms. Thus, $\partial\cdot\Pi=0$, and it follows that $\psi_{p}(x)=\e^{-i\,\mathcal{S}_{p}(x)}$ is indeed an exact solution to the wave equation in the anti-self-dual, partially-depleting flying focus background.

\end{document}